# Accurate Cell Segmentation in Digital Pathology Images via Attention Enforced Networks

Muyi Sun *, KaiqiLi, Yiwen Luo, Xiaoguang Zhou, Zeyi Yao*
School of Automation
Beijing University of Posts and Telecommunications
Beijing, China
{ sunmuyi*, Lkq1997, luoyiwen1995, zxg, yaozeyi*}@bupt.edu.cn

Guanhong Zhang
Institute of Network Technology
Beijing University of Posts and Telecommunications
Beijing, China
zghzgh1779@bupt.edu.cn

*Abstract*—Automatic cell segmentation is an essential step in the pipeline of computer-aided diagnosis (CAD), such as the detection and grading of breast cancer. Accurate segmentation of cells can not only assist the pathologists to make a more precise diagnosis, but also save much time and labor. However, this task suffers from stain variation, cell inhomogeneous intensities, background clutters and cells from different tissues. To address these issues, we propose an Attention Enforced Network (AENet), which is built on spatial attention module and channel attention module, to integrate local features with global dependencies and weight effective channels adaptively. Besides, we introduce a feature fusion branch to bridge high-level and low-level features. Finally, the marker controlled watershed algorithm is applied to post-process the predicted segmentation maps for reducing the fragmented regions. In the test stage, we present an individual color normalization method to deal with the stain variation problem. We evaluate this model on the MoNuSeg dataset. The quantitative comparisons against several prior methods demonstrate the priority of our approach.

*Keywords—cell segmentation, deep learning, attention mechanism, digital pathology images*

## I. Introduction

Cell image analysis plays an important role in automatic pathology diagnosis, as the diagnosis and grading of most diseases depend on the shape and quantity of cells. However, manually segmenting the images by experts is extremely tedious and time-consuming. Thus, it is necessary for researchers to find a way which can free the pathologists and reduce the cost. To this end, convolutional neural network is earning more and more attention with the quick development of computational power. It makes the quantitative analysis possible, and avoids human error at the same time.

Extensive research has been carried out towards automatic cell segmentation. A large number of methods have been introduced, including thresholding [1], region growing algorithm [2], watershed algorithm [3], active contours model [4] and deep learning based method. Here we only name the most widely used ones. Fully convolutional network (FCN) [5], which is a pioneering work of convolutional neural network in image segmentation tasks, emerges in 2014. Similar to FCN, Unet [6] is proposed for medical image segmentation with the structure of encoder-decoder and skip connections. Chen et al. [7] propose an efficient deep contour-aware model (DCAN) under a unified multi-task learning framework. It can not only output accurate probability maps, but also draw clear contours to separate different objects, which makes great progress in cell segmentation. Kumar et al. [2] introduce a deep leaning based network to classify the nuclei and contours, supported by a region growing approach. Naylor et al. [3] employ a deep neural network to separate the nucleus from the background, followed by a watershed algorithm to post-process the probability maps.

All the methods reviewed above have achieved good segmentation results. However, a common limitation exists for them. Most of these methods are based on local features, short of global information, which is adverse to accurate segmentation. In this work, we propose AENet, which builds on the spatial attention module (SAM) and channel attention module (CAM) [8], followed by a feature fusion branch (FFB). The proposed AENet has been confirmed to be effective to perform cell segmentation. Fig.1 shows the procedure of our method. The main contributions are as follows:

(1) We propose a novel network to solve the problem of cell segmentation. In this model, the following modules are contained.

(i) The spatial attention module is employed to obtain non-local information, which can fuse local features and global dependencies adaptively.

(ii) The channel attention module is applied to weight useful feature maps adaptively.

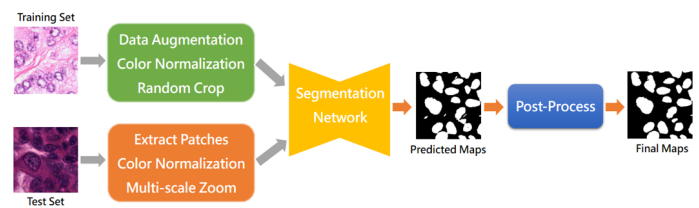

Fig.1. Illustration of the proposed cell segmentation method. In the training stage, data augmentation, color normalization and random crop are implemented on the training set. In the test stage, we apply the patch-based and multi-scale strategy, and the individual color normalization method is applied to deal with stain variation. Finally, we use the marker-controlled watershed algorithm to post-process the predicted maps for reducing the fragment regions.

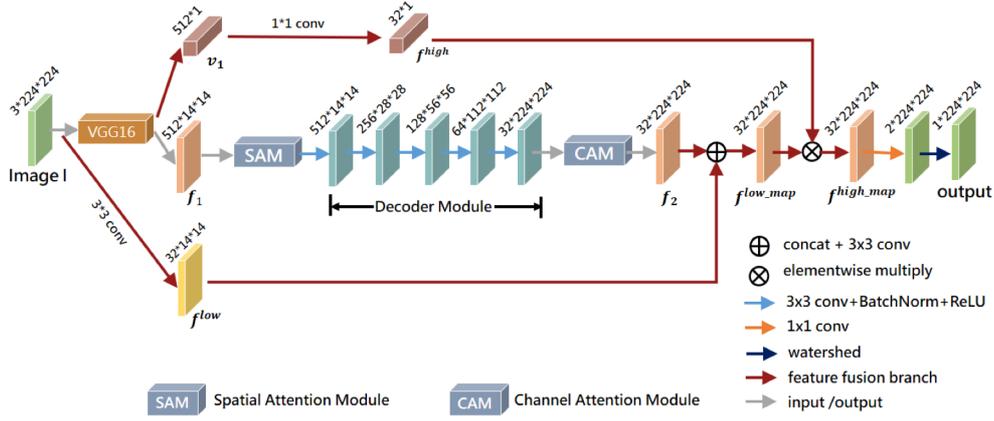

Fig.2. The pipeline of the proposed network. The model consists of six parts: backbone to extract features, spatial attention module, decoder module, channel attention module, feature fusion branch and marker controlled watershed algorithm to post-process the predicted maps. The input image $I \in R^{3*224*224}$ is processed

(iii) We introduce a feature fusion branch to bridge the low-level and high-level features.

(2) We employ the marker-controlled watershed algorithm to post-process the predicted segmentation maps, for reducing the fragmented regions on the prediction maps.

(3) We propose an individual color normalization (ICN) method to deal with the stain variation problem by normalizing the color of pathology images individually.

(4) Extensive experiments on the MoNuSeg dataset and comparisons with prior methods have proved that our method surpasses other segmentation methods.

## II. METHOD

### A. Network Overview

The proposed network utilizes VGG16 [9] as the backbone to extract features. As shown in Fig.2, we take the image $I \in R^{3*224*224}$ as the input of VGG16. The output is a set of feature maps $f_1$ and a feature vector $v_1$. After the feature maps $f_1$ is fed into spatial attention module, decoder module and channel attention module successively, the low-level features $f_{low}$ extracted from the input image I and the high-level features $f_{high}$ obtained from VGG16 will be combined to get more location and global information. In addition, marker controlled watershed algorithm is applied to post-process the predicted semantic segmentation maps aimed at reducing the fragmented regions.

### B. Spatial Attention Module

Recently, a lot of research [10, 11] has proved that the mere use of local features extracted from the traditional fully convolutional networks will result in misclassification. This is because considering only local features rather than global information including contextual semantic information is not sufficient to fulfill complex segmentation tasks. The lack of global information is adverse to accurate segmentation. Therefore, in order to obtain global contextual information, we utilize the spatial attention module to obtain the spatial dependency between any pair of pixels, which uses the self-attention mechanism following the previous work [8]. For a specific position, its feature is equal to the weighted sum of all positions' features. The weight is determined by the similarity between the two pixels, which means the pixels with similar features will promote each other during training, regardless of their distance in the spatial dimension.

As illustrated in Fig.3. For an input feature $A \in R^{C*H*W}$, we feed it into three convolution layers whose kernel size is 1*1 to produce three feature maps $Q, K, V \in R^{C'*H*W}$ before they are reshaped to $Q', K', V' \in R^{C'*N}$, where N is equal to $H*W$. After this, we multiply the transposed Q' by K' so that the result can pass through the softmax layer to obtain the spatial attention map $S \in R^{N*N}$. The formula of softmax is defined as follows:

$$s_{ji} = \frac{exp(B_i \cdot C_j)}{\sum_{i=1}^{N} exp(B_i \cdot C_j)} \quad (1)$$

$S_{ji}$ denotes the effect of the $i^{th}$ pixel on the $j^{th}$ pixel. If the two pixels become more similar, the value of $S_{ji}$ will be higher.

Then the result of multiplying the transpose of S by V' will be reshaped as $S' \in R^{C*H*W}$. The final output F is the element-wise sum of S' and the input feature A.

In [8], the spatial attention module is applied after the decoder module. However, it is computationally expensive. So we add the spatial attention module after the encoder module

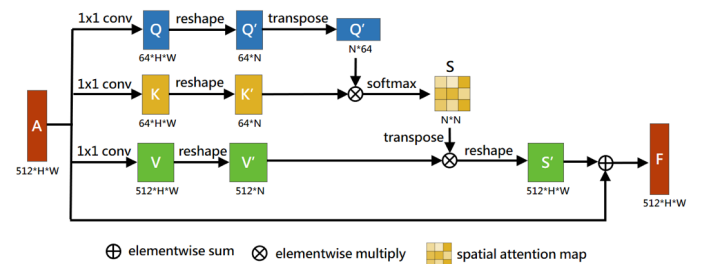

where the output strides = 16, which can greatly reduce the amount of calculation.

Fig.3. Spatial attention module diagram.

## C. Channel Attention Module

Similar to the spatial attention module, channel attention module, which applies the self-attention mechanism to obtain the correlations of any two feature maps, is able to exploit the dependencies between different channels [8]. Each feature map is updated by the sum of all feature maps which are weighted according to the dependency between the two channels.

As shown in Fig.4, we directly reshape the input feature $A \in R^{C*H*W}$ into three identical feature maps Q,K,V$\in R^{C*N}$, where N represents the product of H and W. We then multiply K by the transpose of Q. The result passes through the softmax layer whose formula is interpreted in (2), then the affinities map $S \in R^{C*C}$ is obtained.

$$C_{(x,y)} = \frac{exp(F_{(x)} \cdot F_{(y)})}{\sum_{x=1}^{C} exp(F_{(x)} \cdot F_{(y)})} \quad (2)$$

$C_{(x,y)}$ denotes the correlation between the $x_{th}$ channel and the $y_{th}$ channel. If the two channels' correlation is stronger, the value of $C_{(x,y)}$ will be larger.

We reshape the product of S and V into $S' \in R^{C*H*W}$. At last, we get the final updated feature maps F by sum S' and the input feature maps A.

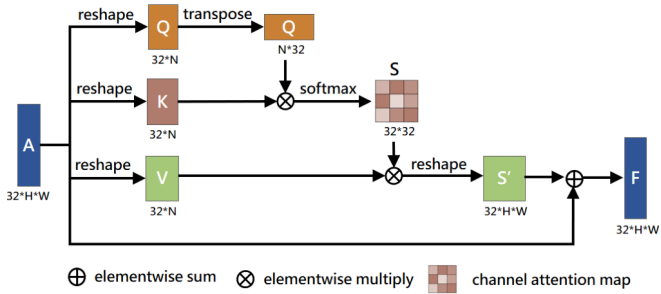

Fig.4. Channel attention module diagram.

## D. Feature Fusion Branch

In this section, we fuse low-level features and high-level features successively. Fig.2 illustrates the forward flow of this section by using solid red arrows.

Generally speaking, the cells are very small, and the regions we need to focus on are usually pixel-level with respect to the cell segmentation task. Therefore, precise location information is essential to segment cells accurately. We obtain rich location information from the low-level feature maps with high resolution, which are extracted from input image I directly. In addition, it is also necessary to fuse global contextual information. To this end, we integrate high-level features produced by VGG16 to get better segmentation results. The details are explained below.

First of all, the feature maps $f^{low}$ is generated from the image I by implementing a 3*3*3*32 convolution, followed by a batch normalization layer and a ReLU layer. Secondly, a convolution layer applies a 512*1*1*32 kernel on $v_1$ to produce $f^{high}$. Thirdly, $\oplus$ denotes concatenation and convolution operation, taking $f^{low}$ and $f_2$ as inputs to produce $f^{low\_map}$. $\otimes$ denotes element-wise multiply between $f^{low\_map}$ and $f^{high}$ to obtain $f^{high\_map}$.

## E. Watershed Algorithm

The results obtained by the deep neural network are encouraging, but a lot of fragmented regions are contained. It will seriously reduce the accuracy of cell segmentation and affect the accurate diagnosis of pathologists. To solve this issue, the marker controlled watershed algorithm is applied to post-process the segmentation results for reducing the noise.

The watershed algorithm is a mathematical morphology method based on topological theory whose principle is to treat the image as topographic geomorphology. The gray value of each pixel represents its altitude. For each local minimum, the boundary of its area of influence is called watershed [3].

As shown in Fig.5, (a) is the original prediction map by the deep neural network. Firstly, we distinguish the part that must be the background area whose pixel value is 0 in (a), and the result is shown in (b). (c) is the distance map after distance transformation, which means calculating the distance between the pixels and their nearest 0-value pixel. As illustrated in (d), a threshold is applied to the distance map to extract the foreground. In (e), the white area indicates the unknown area which is neither foreground nor background, and the borderline will be established in this area. The sub-areas in different colors represent different markers. (f) is the final maps processed by the watershed algorithm.

Fig.5. The pipeline of the watershed algorithm.

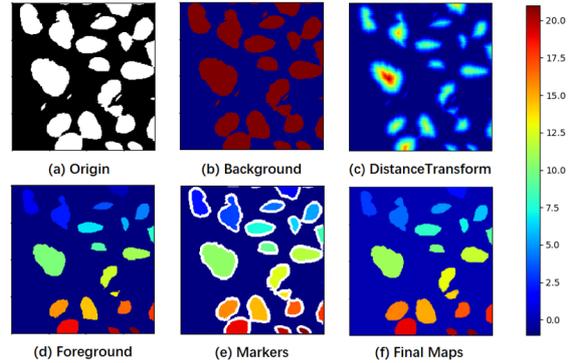

## III. EXPERIMENT

### A. Datasets

We evaluate our method on MoNuSeg. It is initially used in [2], composed of 30 images, each with size 1000*1000. These images are taken from seven organs: 6 of breast, 6 of liver, 6 of kidney, 6 of prostate, 2 of bladder, 2 of colon and 2 of stomach, with annotations of 21623 individual nuclei.

We split the dataset into training set, same organ test set (ST) and different organ test set (DT) in the same fashion as [2]. As shown in Table 1, the training set consists of 16 images, including over 13000 annotated nucleus. They are from four organs: breast, kidney, liver and prostate. The eight images in same organ test set are from the organs which are represented in the training set. And the different organ test set is composed of six images from three organs which not appear in the training set. Thus, the different organ test set requires stronger generalization and robustness, which will be a huge challenge for our model.

TABLE I. MONUSEG DATASET SPLITS (SOURCE FROM [2])

| Data Subset | Breast | Liver | Kidney | Prostate | Bladder | Colon | Stomach |
|---|---|---|---|---|---|---|---|
| Training set | 4 | 4 | 4 | 4 | - | - | - |
| Same organ test set | 2 | 2 | 2 | 2 | - | - | - |
| Different organ test set | - | - | - | - | 2 | 2 | 2 |

### B. Data Preprocess

*1) Data Augmentation*

In this work, some data augmentation strategies are implemented on training sets, including flip, rotation, zoom and random crop. The original 16 training images are augmented to 96 after being horizontally flipped, vertically flipped, and rotated by 90°, 180°, 270° counterclockwise. Then, the 96 images are zoomed with scale s = [0.5, 0.75, 1.0, 1.25, 1.5, 1.75]. During training, we crop 224*224 patches as the input of the network.

*2) Patch-based and Multi-scale Inference*

If the original test images of 1000*1000 are directly fed into the network, it will consume a lot of computing resources and lead to inaccurate segmentation due to the small portion of each single cell in the image. Thus, in order to save computing resources and get more accurate segmentation results, we employ a patch-based strategy.

During testing, we feed patches of 200*200 into the network, and reassemble the patches after inference. In the multi-scale (MS) and flip evaluation, the images are scaled to 0.5, 0.75, 1.0, 1.25, 1.5, 1.75, and 2.0, then horizontally flipped. The prediction maps are up-sampled to the target size using bilinear interpolation and combined by averaging.

*3) Individual Color Normalization*

Moreover, color normalization is employed on both training and testing images. The normalization's formula is defined as follows:

$$x_{normalization} = \frac{x_i - \mu}{\sigma} \quad (3)$$

Where μ represents mean value and σ denotes standard deviation. In particular, in the training phase, μ and σ are obtained by calculating RGB values of the whole training set. However, in the testing phase, we calculate μ and σ of each image individually for the reason that the colors of images in the test set are significantly various. The stain variation makes the segmentation more difficult and has stricter requirements for our approach. Thus, we propose this normalization strategy.

The calculation formula is defined as follows:

$$x_{normalization} = \frac{x_i - u_i}{\sigma_i} \quad (4)$$

Where $u_i$ and $\sigma_i$ denote the mean value and standard deviation of the $i_{th}$ image. This normalization method significantly improves the performance of same organ test set.

### C. Implementation and Training Details

We adopt VGG16 network pre-trained on the ImageNet [12] as our encoder backbone. The output feature maps of the encoder structure are 1/16 size of input image $I \in R^{224*224}$. They are up-sampled to the same size as the input image I by bilinear interpolation to obtain predicted semantic labels of each pixel. We set the maximum epoch as 150. The training algorithm is run until the maximum epoch or convergence. The initial learning rate is 0.0006, which is empirically optimized. During training, we change the learning rate by three policies. In the first 50 epochs, the learning rate is equal to the initial learning rate. It is halved between the 50$^{th}$ and 80$^{th}$ epochs. For the rest of epochs, the poly learning rate policy ( $lr = initial\_lr * (1 - \frac{iter}{total\_iter})^{epoch}$ ) is applied. We choose Adam with default parameters as the optimizer, and set batch size as 32. The experiments are implemented on PyTorch version 1.1.0 with two Nvidia Titan Xp GPUs.

### D. Evaluation Metrics

Cell segmentation task is a binary classification problem, aimed at distinguishing whether each pixel of the input image is cell (class 0) or background (class 1). We suppose that g represents ground truth, and $p$ indicates prediction. By comparing $g$ and p, we can get the following four indexes: TP (count of pixels for which p=0 and g=0), TN (count of pixels for which p=1 and g=1), FP (count of pixels for which p=0 and g=1) and FN (count of pixels for which g=0 and p=1). Based on these four indexes, we evaluate the performance of segmentation in terms of accuracy, recall, precision, F1-score, mean intersection-over-union (mIoU) and dice coefficient.

Accuracy is defined as:

$$Accuracy = \frac{TP+TN}{TP+TN+FP+FN} \quad (5)$$

Recall is defined as:

$$Recall = \frac{TP}{TP+FN} \quad (6)$$

Precision is defined as:

$$Precision = \frac{TP}{TP+FP} \quad (7)$$

F1_score considers both precision and recall. As shown in (8), P denotes precision and R denotes recall.

$$F1\_score = \frac{2*P*R}{P+R} \quad (8)$$

mIoU is defined as:

$$mIoU = \frac{1}{N}\sum_c \frac{TP}{TP+FP+FN} \quad (9)$$

where N is the number of classes. Dice coefficient is defined as:

$$\text{Dice} = \frac{2*TP}{TP+FP+FN} \quad (10)$$

*E. Ablation Study*

We conduct ablation studies to evaluate the contributions of each proposed component to the overall performance of our model. Specifically, we performed the following experiments: (1) contribution of spatial attention module and channel attention module; (2) with and without feature fusion branch; (3) effect of watershed post-process strategy; (4) whether to normalize each image individually during testing; (5) multi-scale inference compared to single-scale inference. The results of the first three experiments are shown in Table II and Table III, where Table II is for same organ test set and Table III is for different organ test set.

*1) Effect of Spatial Attention Module and Channel Attention Module*

As shown in Table II and Table III, CAM significantly improves the performance of both ST and DT. In particular, the huge gain in the dice score of 1.5% for ST and 3.1% for DT proves the channel attention module can effectively improve the segmentation results. SAM has great effects on ST, while it is not so effective on DT. For ST, the value of F1-score, dice and mIoU all have been greatly improved. However, they have hardly increased for DT. The results reveal that CAM and SAM play important roles in the improvement of the performance.

*2) Effect of Feature Fusion Branch and Post-processing*

Apparently, both feature fusion branch and the watershed algorithm can increase the value of F1-score. After applying the feature fusion branch and watershed algorithm, an obvious improvement of F1-score is obtained, with a gain of 1.2% for ST. The score of dice is also increased by 1.2% for DT.

TABLE II. EVALUATION OF THE EFFECTIVENESS OF THE PROPOSED MODULES ON SAME ORGAN TEST SET

| CAM | SAM | FFB | WS | F1-score | Dice | mIoU |
|---|---|---|---|---|---|---|
| | | | | 0.779 | 0.747 | 0.734 |
| √ | | | | 0.797 | 0.762 | 0.749 |
| √ | √ | | | 0.827 | 0.790 | 0.775 |
| √ | √ | √ | | 0.833 | 0.785 | 0.772 |
| √ | √ | √ | √ | 0.839 | 0.792 | 0.779 |

TABLE III. EVALUATION OF THE EFFECTIVENESS OF THE PROPOSED MODULES ON DIFFERENT ORGAN TEST SET

| CAM | SAM | FFB | WS | F1-score | Dice | mIoU |
|---|---|---|---|---|---|---|
| | | | | 0.869 | 0.793 | 0.752 |
| √ | | | | 0.892 | 0.824 | 0.803 |
| √ | √ | | | 0.891 | 0.820 | 0.806 |
| √ | √ | √ | | 0.893 | 0.825 | 0.807 |
| √ | √ | √ | √ | 0.897 | 0.832 | 0.812 |

*3) Effect of Individual Color Normalization and Multi-scale Inference*

In this part, we qualitatively demonstrate the effectiveness of individual color normalization and multi-scale inference. The results of all combinations are shown in Table IV and Table V. For ST, ICN can improve the score of recall greatly but suffers from the low value of precision, while it is the opposite for DT. When ICN and MS are employed at the same time for ST, the score of precision increases by 2.4% compared to using ICN only. With respect to DT, ICN and MS increase the score of precision by 3.1% and 5.6% respectively.

Although the model with individual color normalization shows a slightly lower performance for DT, moderate improvement is obtained for ST. The reason is that the color of images in ST is more various than DT. As shown in Fig.6, a clear improvement is observed by comparing (c) and (d). For darkly stained images, the score of precision is increased dramatically, while for lightly stained images, the score of recall is improved obviously. These results strongly demonstrate that this method of normalization is helpful to deal with stain variation in pathology images.

As a trade of these issues, we apply individual color normalization and multi-scale inference simultaneously for ST, and only apply multi-scale inference for DT.

TABLE IV. ABLATION STUDY FOR INDIVIDUAL COLOR NORMALIZATION AND MULTI-SCALE INFERENCE ON SAME ORGAN TEST SET

| ICN | MS | Recall | Precision | F1-score | Dice |
|---|---|---|---|---|---|
| | | 0.770 | 0.922 | 0.839 | 0.792 |
| | √ | 0.759 | 0.894 | 0.821 | 0.775 |
| √ | | 0.800 | 0.884 | 0.840 | 0.804 |
| √ | √ | 0.787 | 0.908 | 0.843 | 0.812 |

TABLE V. ABLATION STUDY FOR INDIVIDUAL COLOR NORMALIZATION AND MULTI-SCALE INFERENCE ON DIFFERENT ORGAN TEST SET

| ICN | MS | Recall | Precision | F1-score | Dice |
|---|---|---|---|---|---|
| | | 0.832 | 0.832 | 0.832 | 0.812 |
| | √ | 0.810 | 0.888 | 0.847 | 0.817 |
| √ | | 0.798 | 0.863 | 0.829 | 0.798 |
| √ | √ | 0.786 | 0.909 | 0.843 | 0.804 |

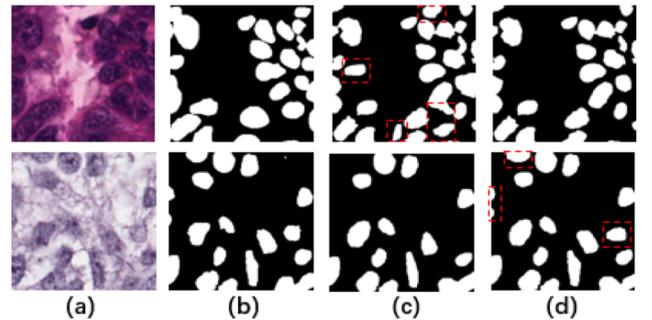

Fig.6. Segmentation results of same organ test set (a) input images (b) ground truth (c) predictions generated by network without individual color normalization and multi-scale inference (d) predictions generated by network with individual color normalization and multi-scale inference

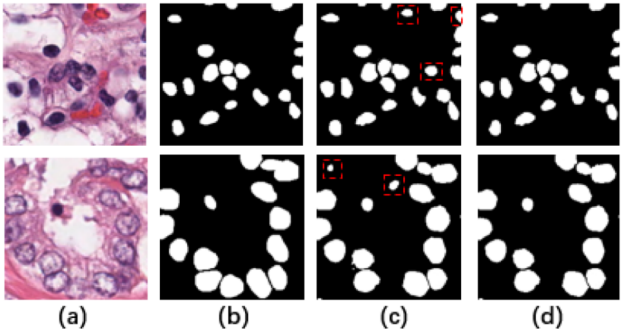

Fig.7. Segmentation results of different organ test set (a) input images (b) ground truth (c) predictions generated by network without multi-scale inference (d) predictions generated by network with multi-scale inference

*4) Comparisons with Other Models*

To make our method more convincing, we compare the proposed network against general segmentation frameworks: Fcn [5], Unet [6], FPN [13], PSPNet [14] and SegNet [15]. Fcn is a widely employed network for segmentation tasks. PSPNet and SegNet are introduced to deal with the task of scene parsing. FPN constructs feature pyramids to detect objects at different scales, which is in favor of cell segmentation task. Compared with these networks, Unet is the first to solve the problem of medical image segmentation and achieves excellent performance.

TABLE VI. QUANTITATIVE COMPARISON AGAINST OTHER METHODS ON SAME ORGAN TEST SET

|  | Accuracy | F1-score | Dice | mIoU |
| --- | --- | --- | --- | --- |
| Fcn | 0.893 | 0.779 | 0.747 | 0.734 |
| Unet | 0.892 | 0.776 | 0.745 | 0.732 |
| FPN | 0.880 | 0.752 | 0.727 | 0.714 |
| PSPNet | 0.816 | 0.636 | 0.616 | 0.615 |
| SegNet | 0.839 | 0.671 | 0.605 | 0.625 |
| ours | 0.921 | 0.843 | 0.812 | 0.787 |

TABLE VII. QUANTITATIVE COMPARISON AGAINST OTHER METHODS ON SAME ORGAN TEST SET

|  | Accuracy | F1-score | Dice | mIoU |
| --- | --- | --- | --- | --- |
| Fcn | 0.869 | 0.793 | 0.752 | 0.720 |
| Unet | 0.855 | 0.764 | 0.745 | 0.705 |
| FPN | 0.844 | 0.749 | 0.716 | 0.682 |
| PSPNet | 0.799 | 0.646 | 0.594 | 0.579 |
| SegNet | 0.882 | 0.814 | 0.777 | 0.743 |
| ours | 0.898 | 0.843 | 0.804 | 0.772 |

As represented in Table VI and Table VII, our method achieves better results than others. PSPNet and SegNet performs poorly on ST, and FPN beats them by a fairly large margin. But for DT, SegNet perfroms well. Notably, Unet outperforms FPN by an increase of 1.8% and 2.9% for ST and DT with respet to the score of dice, respectively. Compared to Unet, the performance of Fcn for ST is qualitatively similar. However, for DT, Fcn achieves an improvement of 2.9% for F1-score. Moreover, for both ST and DT, our method can achieve the highest scores of accuracy, F1-score, dice and mIoU, compared to other methods. It is worth mentioning that our method outperforms Fcn by 2.8% and 2.9% on ST and DT with respect to accuracy respectively.

## IV. CONCLUSION

In this paper, we present a deep convolutional network (AENet) for cell segmentation in multi-tissue pathology images. The proposed network employs two attention modules and a feature fusion branch. Marker controlled watershed algorithm follows to post-process the predicted segmentation maps, which can reduce the noise obviously. In addition, we report the effect of individual color normalization and multi-scale inference. We evaluate our model on two different test sets: same organ test set and different organ test set. In particular, the images in different organ test set are from the tissues which are not represented in training set. The result shows that our approach outperforms other prior methods and demonstrates the ability of our model to generalize well on the images from unseen tissues. This makes the presented approach potentially likely to translate well to a practical setting owing to the strong generalization and robustness.